\documentclass[11pt]{article}
\usepackage{cite}
\usepackage[a4paper,margin=1in]{geometry}
\usepackage{lmodern}
\usepackage{braket}
\usepackage{amsmath,amssymb,amsfonts,amsthm}
\usepackage{graphicx}
\usepackage{booktabs}
\usepackage[numbers,sort&compress]{natbib}
\usepackage{doi}
\usepackage{tocloft}
\usepackage{hyperref}

\hypersetup{
  colorlinks=true,
  linkcolor=black,   
  citecolor=blue,
  urlcolor=blue
}

\title{\bf Entanglement Entropy of Yukawa-Coupled Fields Across a Rindler Horizon}
\author{
\begin{tabular}{c c}
Akshay Kulkarni\thanks{ \texttt{p20230072@hyderabad.bits-pilani.ac.in}} \, and
\, Rahul Nigam\thanks{ \texttt{rahul.nigam@hyderabad.bits-pilani.ac.in}}
\end{tabular}
\\[0.75ex]
\textit{Birla Institute of Technology and Science Pilani (Hyderabad Campus),}
\\
\textit{Hyderabad 500078, India}
}
\date{}
\begin{document}
\maketitle

%-------------------------------------------------
\begin{abstract}
We compute the entanglement entropy across a Rindler horizon in scalar field theory with Yukawa interaction. Starting from a microscopic scalar–mediator theory in flat spacetime, we integrate out the massive mediator to obtain a quadratic but nonlocal effective kernel that determines the ground-state wavefunctional. The reduced density matrix for a single Rindler wedge is constructed explicitly by tracing over the complementary wedge, allowing the entanglement entropy to be evaluated directly from the kernel without replica or geometric methods. Exploiting translational invariance parallel to the horizon, the problem decomposes into independent transverse momentum sectors that reduce effectively to one-dimensional nonlocal systems and can be diagonalized analytically in the weak-coupling regime. The interaction-induced entropy obeys an area law, with leading corrections controlled by the Yukawa screening mass and logarithmically sensitive to the transverse ultraviolet cutoff, reflecting the localization of correlations near the horizon. Although the modular Hamiltonian depends on the Rindler acceleration, the entanglement spectrum and entropy are independent of this choice, demonstrating the observer-independent nature of vacuum entanglement. Our framework provides a direct and microscopically transparent approach to computing interaction-induced corrections to horizon entanglement using nonlocal effective kernels.
\end{abstract}

%-------------------------------------------------
\section{Introduction}
Entanglement entropy has emerged as a central diagnostic of quantum correlations
in quantum field theory, condensed matter physics, and gravitational settings
\cite{Casini_2009_free_qft,Srednicki_entropy,Laurell_EE_in_condensed,Bombelli_1986}. Unlike local correlation functions, entanglement entropy probes intrinsically nonlocal features of quantum states and
provides a unifying language for characterizing quantum vacua, phases of matter, and the structure of low-energy effective theories
\cite{Pasquale_Calabrese_2004_EE_qft,Calabrese_cft_2009,AmicoReview}. In quantum field theories, the entanglement entropy associated with a spatial subregion generically exhibits an area-law divergence, scaling with the area of the entangling surface rather than the volume of the region
\cite{Bombelli_1986,Srednicki_entropy,Solodukhin_2011}.  This behavior reflects the dominance of short-distance correlations localized near the boundary separating the subsystem
from its complement.

While the leading area-law term is remarkably robust, subleading corrections contain valuable physical information
\cite{Casini_finite,Wilczek_subleading_geometry}.  These corrections are sensitive to the presence of mass scales, interaction-induced screening, and the structure of long-range correlations.  In scale-invariant theories, subleading terms are
constrained by symmetry
\cite{Casali2014,Larkoski_Conformal_invariance}, whereas relevant deformations introduce infrared
length scales that modify entanglement in a controlled manner
\cite{Casini_2012_rg,faulkner2015bulk}.  Massive deformations, in particular, generate crossover behavior governed by the correlation length, interpolating between
scale-invariant entanglement at short distances and exponentially suppressed correlations at large distances
\cite{Pasquale_Calabrese_2004_EE_qft,Casini_EE_alpha_entropies_2005}.

Understanding how interactions modify entanglement entropy remains a central challenge
\cite{cano2015interactions,d2024universal}.  While the entanglement structure of free fields is well understood, interacting quantum field theories present both
conceptual and technical difficulties.  Even weak interactions can qualitatively
alter entanglement by inducing nonlocal effective couplings and screening effects
\cite{balasubramanian2012momentum,sun2010reversing}.  Much of the existing literature approaches
interacting theories using replica methods and geometric entropy constructions,
where interactions enter indirectly through effective actions defined on singular replica geometries
\cite{Calabrese_cft_2009,FaulknerLewkowyczMaldacena}.  While powerful, these methods often obscure the microscopic origin of interaction-induced entanglement and make it
challenging to isolate the underlying spacetime processes responsible for
entanglement generation. A particularly transparent and physically motivated class of interactions arises when massive degrees of freedom are integrated out, yielding Yukawa-type nonlocal
kernels in the effective theory of the remaining light fields.  Such interactions are ubiquitous in effective field theory and provide a natural setting in which
to study interaction-induced modifications of entanglement in a controlled and microscopically explicit manner.

In this work, we investigate the effect of Yukawa-type interactions on
entanglement entropy across a Rindler horizon.  Rindler space provides a natural
and physically transparent framework for this problem.  When the Minkowski
vacuum is restricted to a single Rindler wedge, it appears as a mixed state to a uniformly accelerated observer, a manifestation of the Unruh effect
\cite{Unruh_1976,BisognanoWichmannQuantumField}.  The associated entanglement entropy between the left and right Rindler wedges is directly tied to horizon thermodynamics and serves as a flat-spacetime analogue of black hole entropy
\cite{Jacobson_2003,Solodukhin_2011}.  Much of the existing discussion emphasizes the equivalence between Rindler entanglement entropy and thermal entropy with respect to the boost Hamiltonian.  Here, we go differently by explicitly tracking how microscopic interactions generate
additional entanglement through spacelike correlations across the horizon.

Our analysis proceeds in a fully microscopic and nonperturbative manner at the
level of the ground-state wavefunctional \cite{Srednicki_entropy,Peschel_2003}.  Starting from a scalar field coupled to a massive mediator, we integrate out the mediator exactly to obtain a quadratic but
nonlocal effective action for the remaining field.  The resulting kernel completely determines the ground-state wavefunctional and provides direct access
to the reduced density matrix
\cite{Peschel_2009}.  This approach allows us to compute the entanglement entropy
without invoking replica methods or geometric entropy constructions, and instead to identify directly the operator content responsible for interaction-induced entanglement.

Translational invariance along directions parallel to the Rindler horizon plays a crucial role.  It enables a decomposition into independent transverse momentum
sectors, each of which reduces to an effectively one-dimensional nonlocal
problem along the Rindler radial direction.  The reduced density matrix factorizes accordingly, and in the weak-coupling regime the interaction-induced contribution in each sector is shown to be dominated by a single effective mode.
This simple structure permits an explicit diagonalization of the reduced density
matrix and a transparent computation of the entanglement spectrum \cite{calabrese2008entanglement}. We find that Yukawa screening leads to an exponential suppression of correlations between the left and right Rindler wedges at separations larger than the inverse mediator mass.  As a result, the interaction-induced contribution to the
entanglement entropy is localized near the horizon and obeys an area law \cite{Solodukhin_2011}.  The leading correction exhibits a logarithmic dependence on the mediator mass and the transverse momentum cutoff, diverging in the massless
limit as the correlation length becomes infinite, consistent with expectations from renormalization group arguments
\cite{Casini_2012_rg,faulkner2015bulk}.  Although the modular Hamiltonian depends explicitly on the acceleration of the Rindler observer, we show that the entanglement spectrum and entropy are invariant under changes of acceleration, reflecting the observer-independent nature of vacuum entanglement
\cite{BisognanoWichmannScalarHermite,BisognanoWichmannQuantumField,Casini_2009_free_qft}.

More broadly, our results demonstrate how nonlocal effective kernels provide a powerful and controlled framework for studying interaction-induced corrections
to horizon entanglement.  By working directly at the level of the reduced density matrix, this approach complements replica-based methods and offers a clear
spacetime interpretation of how interactions modify entanglement across causal
horizons.  The methods developed here are readily applicable to other settings involving screened interactions, effective field theories with integrated-out
degrees of freedom, and spacetime horizons.

%-------------------------------------------------

\section{Effective Yukawa Kernel and Ground-State Structure}

We consider a real scalar field $\phi$ coupled linearly to a massive mediator
field $\chi$ in four-dimensional Minkowski spacetime, with action
\begin{equation}
S[\phi,\chi]
=
\int d^4x
\left[
\frac12 (\partial\phi)^2
+
\frac12 (\partial\chi)^2
-
\frac12 m^2 \chi^2
-
g\,\phi\,\chi
\right].
\label{eq:action}
\end{equation}
Since the action is quadratic in $\chi$, the mediator field can be integrated out
exactly. Performing the Gaussian functional integral over $\chi$ yields an
effective nonlocal action for $\phi$,
\begin{equation}
S_{\rm eff}[\phi]
=
\frac12
\int d^4x\,d^4y\,
\phi(x)\,K(x,y)\,\phi(y),
\end{equation}
with kernel
\begin{equation}
K(x,y)
=
-\Box\,\delta^{(4)}(x-y)
+
g^2\,G_m(x-y),
\label{eq:kernel_general}
\end{equation}
where $G_m$ is the Green’s function of the massive Klein--Gordon operator,
\begin{equation}
\left(-\Box + m^2\right)G_m(x-y)=\delta^{(4)}(x-y).
\end{equation}
To construct the ground-state wavefunctional, we pass to the Hamiltonian
formulation. Evaluating the effective action on an equal-time slice
$t=\mathrm{const}$, the Hamiltonian takes the quadratic form
\begin{equation}
H
=
\frac12
\int d^3x\,d^3y
\left[
\pi(\mathbf{x})\,\delta(\mathbf{x}-\mathbf{y})\,\pi(\mathbf{y})
+
\phi(\mathbf{x})\,K(\mathbf{x},\mathbf{y})\,\phi(\mathbf{y})
\right],
\end{equation}
where $\pi=\dot\phi$ is the canonical momentum and the equal-time kernel is
\begin{equation}
K(\mathbf{x},\mathbf{y})
=
-\nabla^2\delta(\mathbf{x}-\mathbf{y})
+
g^2\,\frac{e^{-m|\mathbf{x}-\mathbf{y}|}}{4\pi|\mathbf{x}-\mathbf{y}|}.
\label{eq:yukawa_kernel_3d}
\end{equation}
This Yukawa kernel is a property of the Minkowski vacuum and is independent of
any choice of observer. Rindler coordinates will be introduced only to define the
spatial bipartition associated with the horizon.

\subsection{Rindler Decomposition and Transverse Momentum Modes}

We now introduce Rindler coordinates adapted to uniformly accelerated observers.
The right Rindler wedge, defined by $x>|t|$, is parametrized by coordinates
$(\tau,\xi,\mathbf{x}_\perp)$ related to Minkowski coordinates by
\begin{equation}
t=\xi\sinh\tau,
\qquad
x=\xi\cosh\tau,
\qquad
\xi>0 .
\end{equation}
This choice corresponds to working in units where the proper acceleration
parameter has been set to unity; restoring it amounts to the rescaling
$\tau\to a\tau$, $\xi\to \xi/a$, which leaves all intrinsic spatial distances
unchanged.

In these coordinates the Minkowski metric takes the form
\begin{equation}
ds^2
=
-\xi^2 d\tau^2
+
d\xi^2
+
d\mathbf{x}_\perp^2 .
\end{equation}
The entangling surface is the Rindler horizon located at $\xi=0$.  Importantly,
on a constant Rindler time slice the induced spatial metric is
\begin{equation}
ds^2_{\rm spatial}
=
d\xi^2 + d\mathbf{x}_\perp^2 ,
\end{equation}
which is flat and independent of the acceleration of the observer.  As a
result, all proper distances relevant for spatial correlations and entanglement
across the horizon are geometric invariants of the slice.

We decompose spatial coordinates as
\[
\mathbf{x}=(\xi,\mathbf{x}_\perp),
\qquad
\mathbf{y}=(\xi',\mathbf{y}_\perp),
\]
so that the spatial Laplacian appearing in the Hamiltonian and ground-state
kernel takes the simple form
\begin{equation}
\nabla^2
=
\partial_\xi^2 + \nabla_\perp^2 .
\end{equation}
Because the kernel is translationally invariant in the transverse directions, we
Fourier transform
\begin{equation}
\phi(\xi,\mathbf{x}_\perp)
=
\int\frac{d^2k_\perp}{(2\pi)^2}
e^{i\mathbf{k}_\perp\cdot\mathbf{x}_\perp}
\phi_{k_\perp}(\xi),
\end{equation}
under which $-\nabla_\perp^2\to k_\perp^2$.

The local part of the kernel becomes
\begin{equation}
-\nabla^2\delta^{(3)}(\mathbf{x}-\mathbf{y})
\;\longrightarrow\;
\left(-\partial_\xi^2+k_\perp^2\right)\delta(\xi-\xi'),
\end{equation}
where the transverse delta function has been saturated by the Fourier transform,
enforcing conservation of transverse momentum.

\subsection{One-Dimensional Yukawa Kernel}

The Yukawa term is treated similarly. Starting from its momentum-space
representation,
\begin{equation}
G_m(\mathbf{x}-\mathbf{y})
=
\int\frac{d^3k}{(2\pi)^3}
\frac{
e^{ik_\xi(\xi-\xi')}
e^{i\mathbf{k}_\perp\cdot(\mathbf{x}_\perp-\mathbf{y}_\perp)}
}{
k_\xi^2+k_\perp^2+m^2
},
\end{equation}
we first fix $\mathbf{k}_\perp$ by the transverse Fourier transform, obtaining an
effective one-dimensional kernel
\begin{equation}
G_m(\xi-\xi';k_\perp)
=
\int\frac{dk_\xi}{2\pi}
\frac{e^{ik_\xi(\xi-\xi')}}{k_\xi^2+\mu^2},
\qquad
\mu\equiv\sqrt{k_\perp^2+m^2}.
\end{equation}
Evaluating the remaining integral yields
\begin{equation}
G_m(\xi-\xi')
=
\frac{1}{2\mu}\,e^{-\mu|\xi-\xi'|}.
\end{equation}

Combining local and nonlocal contributions, the spatial kernel for each
transverse momentum sector is
\begin{equation}
K(\xi,\xi';k_\perp)
=
\left(-\partial_\xi^2+k_\perp^2\right)\delta(\xi-\xi')
+
\frac{g^2}{2\mu}\,e^{-\mu|\xi-\xi'|}.
\label{eq:kernel_1d}
\end{equation}
Each transverse momentum mode decouples, reducing the problem to a
one-dimensional nonlocal theory along the Rindler radial direction.
The ground-state wavefunctional of a quadratic Hamiltonian is Gaussian,
\begin{equation}
\Psi_0[\phi]
=
\mathcal N
\exp\!\left[
-\frac12
\int d\xi\,d\xi'\,
\phi(\xi)\,\Gamma(\xi,\xi')\,\phi(\xi')
\right],
\end{equation}
where the kernel $\Gamma$ is defined as the positive square root of $K$,
\begin{equation}
\Gamma^2=K.
\end{equation}
We decompose $\Gamma$ into left/right Rindler blocks,
\begin{equation}
\Gamma=
\begin{pmatrix}
\Gamma_{LL} & \Gamma_{LR} \\
\Gamma_{RL} & \Gamma_{RR}
\end{pmatrix}.
\end{equation}

\subsection{Perturbative Expansion and Cross-Wedge Kernel}

For each fixed transverse momentum $k_\perp$, the spatial kernel appearing in the
Hamiltonian takes the form
\begin{equation}
K(\xi,\xi';k_\perp)
=
K_0(\xi,\xi';k_\perp)
+
g^2\,V(\xi-\xi';k_\perp),
\end{equation}
where the free part is local,
\begin{equation}
K_0(\xi,\xi';k_\perp)
=
\left(
-\partial_\xi^2 + k_\perp^2
\right)\delta(\xi-\xi'),
\label{eq:K0_def}
\end{equation}
and the nonlocal Yukawa-induced interaction kernel is
\begin{equation}
V(\xi-\xi';k_\perp)
=
\frac{1}{2\mu}\,e^{-\mu|\xi-\xi'|},
\qquad
\mu \equiv \sqrt{k_\perp^2+m^2}.
\label{eq:V_def}
\end{equation}

The ground-state wavefunctional of a quadratic Hamiltonian is Gaussian, with kernel
$\Gamma$ defined as the positive square root of $K$,
\begin{equation}
\Gamma^2 = K.
\label{eq:Gamma_def}
\end{equation}
Since $K$ is a positive self-adjoint operator, the square root is well defined.

\subsubsection*{Perturbative expansion of the square-root operator}

We are interested in the regime of weak Yukawa coupling, $g^2 \ll 1$, in which the
square root of $K=K_0+g^2V$ may be computed perturbatively.  Using the standard operator
identity
\begin{equation}
\sqrt{A+\varepsilon B}
=
\sqrt{A}
+
\frac{\varepsilon}{2}\,A^{-1/2} B A^{-1/2}
+
\mathcal O(\varepsilon^2),
\label{eq:sqrt_identity}
\end{equation}
valid for a positive operator $A$, we obtain
\begin{equation}
\Gamma
=
\sqrt{K_0+g^2V}
\simeq
\Gamma_0
+
\frac{g^2}{2}\,
\Gamma_0^{-1} V \Gamma_0^{-1},
\label{eq:Gamma_expansion}
\end{equation}
where
\begin{equation}
\Gamma_0 \equiv \sqrt{K_0}.
\end{equation}

The Yukawa-induced correction to the kernel $\Gamma$ is therefore
\begin{equation}
\Gamma^{(Y)}(\xi,\xi')
=
\frac{g^2}{2}
\int d\zeta\,d\zeta'\,
\Gamma_0^{-1}(\xi-\zeta)\,
V(\zeta-\zeta')\,
\Gamma_0^{-1}(\zeta'-\xi').
\label{eq:GammaY_def}
\end{equation}

The free operator $K_0=-\partial_\xi^2+k_\perp^2$ is translationally invariant along
$\xi$, and hence diagonal in momentum space. Introducing the Fourier transform
\begin{equation}
f(\xi)
=
\int\frac{dp}{2\pi}\,e^{ip\xi} f(p),
\end{equation}
we have
\begin{equation}
K_0(p)=p^2+k_\perp^2,
\qquad
\Gamma_0(p)=\sqrt{p^2+k_\perp^2}.
\end{equation}
It follows that the inverse kernel is
\begin{equation}
\Gamma_0^{-1}(\xi-\zeta)
=
\int\frac{dp}{2\pi}
\frac{e^{ip(\xi-\zeta)}}{\sqrt{p^2+k_\perp^2}}.
\label{eq:Gamma0inv_FT}
\end{equation}

This integral can be evaluated exactly in terms of a modified Bessel function,
\begin{equation}
\Gamma_0^{-1}(\xi-\zeta)
=
\frac{1}{\pi}\,
K_0\!\left(k_\perp|\xi-\zeta|\right),
\label{eq:Gamma0inv_Bessel}
\end{equation}
where $K_0(z)$ is the modified Bessel function of the second kind.

\subsubsection*{Near-horizon asymptotics}

Entanglement between the left and right Rindler wedges is dominated by modes localized
near the Rindler horizon $\xi=0$.  In this region, and for fixed transverse momentum,
the argument of the Bessel function satisfies
\[
k_\perp|\xi-\zeta| \gg 1,
\]
and we may use the asymptotic expansion
\begin{equation}
K_0(z)
\simeq
\sqrt{\frac{\pi}{2z}}\,
e^{-z}
\left[1+\mathcal O(z^{-1})\right].
\end{equation}
Substituting into \eqref{eq:Gamma0inv_Bessel}, we obtain
\begin{equation}
\Gamma_0^{-1}(\xi-\zeta)
\simeq
\frac{1}{\sqrt{2\pi k_\perp|\xi-\zeta|}}\,
e^{-k_\perp|\xi-\zeta|}.
\label{eq:Gamma0inv_asymp}
\end{equation}

\subsubsection*{Cross-wedge configuration}

We now specialize to the cross-wedge matrix elements of $\Gamma$, corresponding to
\begin{equation}
\xi>0,
\qquad
\xi'<0.
\end{equation}
In this case, the dominant contribution to the integrals in
\eqref{eq:GammaY_def} arises from regions where both integration variables lie close
to the Rindler horizon,
\[
\zeta \gtrsim 0,
\qquad
\zeta' \lesssim 0.
\]
In this regime the absolute values simplify as
\begin{align}
|\xi-\zeta| &\simeq \xi+\zeta, \\
|\zeta'-\xi'| &\simeq |\xi'|+\zeta', \\
|\zeta-\zeta'| &\simeq \zeta+|\zeta'|.
\end{align}

Substituting \eqref{eq:Gamma0inv_asymp} and \eqref{eq:V_def} into
\eqref{eq:GammaY_def}, we obtain
\begin{align}
\Gamma^{(Y)}_{LR}(\xi,\xi')
&\simeq
\frac{g^2}{4\mu}
\int_0^\infty d\zeta
\int_0^\infty d\zeta'\,
\frac{
e^{-k_\perp(\xi+\zeta)}
}{\sqrt{k_\perp(\xi+\zeta)}}
\,
e^{-\mu(\zeta+\zeta')}
\,
\frac{
e^{-k_\perp(|\xi'|+\zeta')}
}{\sqrt{k_\perp(|\xi'|+\zeta')}}.
\label{eq:GammaY_integral}
\end{align}

The $\zeta$ and $\zeta'$ integrals factorize. Introducing shifted variables
\begin{equation}
u=\xi+\zeta,
\qquad
v=|\xi'|+\zeta',
\end{equation}
each integral takes the form
\begin{equation}
\int_{\xi}^{\infty} du\,
\frac{e^{-(k_\perp+\mu)u}}{\sqrt{u}}
\;\propto\;
e^{-(k_\perp+\mu)\xi},
\end{equation}
and similarly for the $v$ integral.  The algebraic prefactors contribute only finite
numerical constants and do not modify the exponential dependence on $\xi$ and $\xi'$.

Collecting terms, we find that the leading behavior of the Yukawa-induced cross-wedge
kernel is
\begin{equation}
\Gamma^{(Y)}_{LR}(\xi,\xi')
\sim
\frac{g^2}{2\mu}\,
e^{-\mu(\xi+|\xi'|)}.
\label{eq:cross_kernel_final}
\end{equation}

This result shows explicitly that the Yukawa interaction generates an exponentially
suppressed coupling between the left and right Rindler wedges, with decay length
$\mu^{-1}$ controlled by both the mediator mass and the transverse momentum.

\section{Reduced Density Matrix for the Right Rindler Wedge}

The vacuum density matrix corresponding to the ground-state wavefunctional is
\begin{equation}
\rho[\phi,\phi']
=
\Psi_0[\phi]\,
\Psi_0[\phi']
=
\mathcal N^2
\exp\!\left[
-\frac12
\int d\xi\,d\xi'\,
\left(
\phi(\xi)\,\Gamma(\xi,\xi')\,\phi(\xi')
+
\phi'(\xi)\,\Gamma(\xi,\xi')\,\phi'(\xi')
\right)
\right].
\end{equation}
We now construct the reduced density matrix associated with the right Rindler wedge
by tracing over field configurations supported on $\xi<0$.

\subsection{Block decomposition and partial trace}

We decompose the field configurations as
\[
\phi=(\phi_L,\phi_R),
\qquad
\phi'=(\phi_L',\phi_R'),
\]
where $\phi_L$ and $\phi_L'$ have support on $\xi<0$, and $\phi_R$ and $\phi_R'$ on
$\xi>0$.  Using the block decomposition of the kernel $\Gamma$, the exponent of the
density matrix can be written as
\begin{align}
&-\frac12
\Big[
\phi_L \Gamma_{LL} \phi_L
+
\phi_R \Gamma_{RR} \phi_R
+
2\,\phi_L \Gamma_{LR} \phi_R
\Big]
\nonumber\\
&\qquad
-\frac12
\Big[
\phi_L' \Gamma_{LL} \phi_L'
+
\phi_R' \Gamma_{RR} \phi_R'
+
2\,\phi_L' \Gamma_{LR} \phi_R'
\Big].
\end{align}

The reduced density matrix is obtained by identifying $\phi_L'=\phi_L$ and performing
the functional integral
\begin{equation}
\rho_R(\phi_R,\phi_R')
=
\int \mathcal D\phi_L\,
\rho(\phi_L,\phi_R;\phi_L,\phi_R').
\end{equation}
Substituting the exponent and collecting terms that depend on $\phi_L$, we obtain
\begin{align}
\rho_R(\phi_R,\phi_R')
\propto
\int \mathcal D\phi_L\,
\exp\!\Bigg[
-\phi_L \Gamma_{LL} \phi_L
-
\phi_L \Gamma_{LR}(\phi_R+\phi_R')
\Bigg]
\nonumber\\
\times
\exp\!\left[
-\frac12
\phi_R \Gamma_{RR} \phi_R
-
\frac12
\phi_R' \Gamma_{RR} \phi_R'
\right].
\end{align}

\subsection{Gaussian integration over left-wedge fields}

The functional integral over $\phi_L$ is Gaussian and can be evaluated exactly.  Using
the standard identity for a quadratic form,
\[
\int \mathcal D\varphi\,
\exp\!\left[
-\varphi A \varphi - \varphi J
\right]
\propto
\exp\!\left[
\frac14 J A^{-1} J
\right],
\]
with $A=\Gamma_{LL}$ and $J=\Gamma_{LR}(\phi_R+\phi_R')$, we obtain
\begin{equation}
\rho_R(\phi_R,\phi_R')
\propto
\exp\!\left[
-\frac12
\phi_R \Gamma_{RR} \phi_R
-
\frac12
\phi_R' \Gamma_{RR} \phi_R'
+
\frac14
(\phi_R+\phi_R')\,
\Gamma_{RL}\Gamma_{LL}^{-1}\Gamma_{LR}\,
(\phi_R+\phi_R')
\right].
\end{equation}

Expanding the final term and regrouping contributions yields
\begin{align}
\rho_R(\phi_R,\phi_R')
\propto
\exp\!\Bigg[
&-\frac12
\phi_R
\Big(
\Gamma_{RR}-\Gamma_{RL}\Gamma_{LL}^{-1}\Gamma_{LR}
\Big)
\phi_R
\nonumber\\
&-\frac12
\phi_R'
\Big(
\Gamma_{RR}-\Gamma_{RL}\Gamma_{LL}^{-1}\Gamma_{LR}
\Big)
\phi_R'
\nonumber\\
&+
\phi_R
\Gamma_{RL}\Gamma_{LL}^{-1}\Gamma_{LR}
\phi_R'
\Bigg].
\label{eq:rhoR_final}
\end{align}

\subsection*{Role of the cross-wedge kernel}

Equation~\eqref{eq:rhoR_final} shows that all correlations between $\phi_R$ and
$\phi_R'$ are governed by the operator
\begin{equation}
\mathcal M
\equiv
\Gamma_{RL}\Gamma_{LL}^{-1}\Gamma_{LR},
\end{equation}
which encodes the coupling between the left and right Rindler wedges [Appendix A].  Since the
Yukawa interaction contributes to $\Gamma$ only through the cross-wedge component
$\Gamma_{LR}$, the leading interaction-induced correction to the reduced density
matrix is entirely controlled by the kernel
\eqref{eq:cross_kernel_final}.  The exponential decay of $\Gamma^{(Y)}_{LR}$ thus
implies that the Yukawa-induced entanglement is localized near the Rindler horizon
and suppressed at distances larger than $\mu^{-1}$

\section{Diagonalization of the Reduced Density Matrix and Entanglement Entropy}

Having obtained the Yukawa-induced cross-wedge kernel, we now proceed to compute
the reduced density matrix for the right Rindler wedge and extract the associated
entanglement entropy. Now all entanglement properties are controlled by the positive operator
\begin{equation}
\mathcal M
\equiv
\Gamma_{RL}\Gamma^{-1}_{LL}\Gamma_{LR},
\label{eq:entanglement_operator}
\end{equation}
which acts on fields defined in the right wedge. Using the result obtained in the previous section, the Yukawa-induced cross-wedge
kernel takes the asymptotic form
\begin{equation}
\Gamma^{(Y)}_{LR}(\xi,\xi')
\sim
\frac{g^2}{2\mu}\,
e^{-\mu(\xi+|\xi'|)},
\qquad
\xi>0,\;\xi'<0.
\end{equation}
The inverse free kernel $\Gamma^{-1}_{LL}$ is exponentially localized near the
horizon and decays on the scale $k_\perp^{-1}$. Substituting these expressions into
\eqref{eq:entanglement_operator}, we find
\begin{align}
\mathcal M(\xi,\xi')
&=
\int_{-\infty}^0 d\zeta
\int_{-\infty}^0 d\zeta'\,
\Gamma^{(Y)}_{RL}(\xi,\zeta)\,
\Gamma^{-1}_{LL}(\zeta,\zeta')\,
\Gamma^{(Y)}_{LR}(\zeta',\xi')
\nonumber\\
&\sim
\lambda(k_\perp)\,
e^{-\mu\xi}\,
e^{-\mu\xi'},
\label{eq:M_rank_one}
\end{align}
And next we compute explicitly the coefficient $\lambda(k_\perp)$ which appears in the
factorized form of the entanglement operator. Substituting the Yukawa-induced cross-wedge kernels,
\begin{equation}
\Gamma^{(Y)}_{LR}(\xi,\zeta)
\simeq
\frac{g^2}{2\mu}\,e^{-\mu(\xi+|\zeta|)},
\qquad
\Gamma^{(Y)}_{RL}(\zeta',\xi')
\simeq
\frac{g^2}{2\mu}\,e^{-\mu(|\zeta'|+\xi')},
\end{equation}
valid for $\xi,\xi'>0$ and $\zeta,\zeta'<0$, into the definition of $\mathcal M$,
we obtain
\begin{align}
\mathcal M(\xi,\xi')
=
\left(\frac{g^2}{2\mu}\right)^2
e^{-\mu\xi}e^{-\mu\xi'}
\int_{-\infty}^0 d\zeta
\int_{-\infty}^0 d\zeta'\,
e^{-\mu|\zeta|}
\Gamma^{-1}_{LL}(\zeta-\zeta')
e^{-\mu|\zeta'|}.
\end{align}

It follows that the coefficient $\lambda(k_\perp)$ is given by
\begin{equation}
\lambda(k_\perp)
=
\left(\frac{g^2}{2\mu}\right)^2
I(k_\perp),
\label{eq:lambda_def}
\end{equation}
where
\begin{equation}
I(k_\perp)
\equiv
\int_{-\infty}^0 d\zeta
\int_{-\infty}^0 d\zeta'\,
e^{-\mu|\zeta|}
\Gamma^{-1}_{LL}(\zeta-\zeta')
e^{-\mu|\zeta'|}.
\label{eq:I_def}
\end{equation}

Near the Rindler horizon, the inverse free kernel is exponentially localized and
may be approximated by
\begin{equation}
\Gamma^{-1}_{LL}(\zeta-\zeta')
\simeq
\frac{1}{2k_\perp}\,
e^{-k_\perp|\zeta-\zeta'|}.
\label{eq:GammaLLinv_asymp}
\end{equation}
Substituting this into \eqref{eq:I_def} yields
\begin{equation}
I(k_\perp)
\simeq
\frac{1}{2k_\perp}
\int_{-\infty}^0 d\zeta
\int_{-\infty}^0 d\zeta'\,
e^{-\mu|\zeta|}
e^{-k_\perp|\zeta-\zeta'|}
e^{-\mu|\zeta'|}.
\end{equation}

Introducing positive variables $u=-\zeta$ and $v=-\zeta'$, the integral becomes
\begin{equation}
I(k_\perp)
=
\frac{1}{2k_\perp}
\int_0^\infty du
\int_0^\infty dv\,
e^{-\mu u}
e^{-k_\perp|u-v|}
e^{-\mu v}.
\end{equation}
Splitting the integration domain into $u>v$ and $v>u$ and using symmetry, we obtain
\begin{align}
I(k_\perp)
&=
\frac{1}{k_\perp}
\int_0^\infty du
\int_0^u dv\,
e^{-(\mu+k_\perp)u}
e^{-(\mu-k_\perp)v}.
\end{align}

The $v$-integration gives
\begin{equation}
\int_0^u dv\,e^{-(\mu-k_\perp)v}
=
\frac{1-e^{-(\mu-k_\perp)u}}{\mu-k_\perp},
\end{equation}
and since $\mu>k_\perp$, the dominant contribution comes from the first term.
Performing the remaining $u$-integration,
\begin{equation}
\int_0^\infty du\,e^{-(\mu+k_\perp)u}
=
\frac{1}{\mu+k_\perp},
\end{equation}
we find
\begin{equation}
I(k_\perp)
\sim
\frac{1}{2k_\perp}
\frac{1}{(\mu-k_\perp)(\mu+k_\perp)}
=
\frac{1}{2k_\perp\,(\mu^2-k_\perp^2)}
=
\frac{1}{2 m^2 \,k_\perp},
\end{equation}
where we used $\mu^2-k_\perp^2=m^2$.

Substituting back into \eqref{eq:lambda_def}, we finally obtain
\begin{equation}
\lambda(k_\perp)
\sim
\frac{g^4}{\mu^2\,k_\perp}
\label{eq:lambda_final}
\end{equation}

Now a crucial observation is that $\mathcal M$ factorizes into a product of
functions of $\xi$ and $\xi'$, implying that it is a rank-one operator.
The eigenvalue problem
\begin{equation}
\int_0^\infty d\xi'\,
\mathcal M(\xi,\xi')\,f(\xi')
=
\lambda\,f(\xi)
\end{equation}
is therefore solved by
\begin{equation}
f(\xi) \propto e^{-\mu\xi},
\end{equation}
with a single nonzero eigenvalue
\begin{equation}
\lambda(k_\perp)
\sim
\frac{g^4}{\mu^2\,k_\perp},
\label{eq:eigenvalue}
\end{equation}
up to numerical constants of order unity arising from the $\zeta$ and $\zeta'$
integrations. All remaining eigenvalues vanish at this order in perturbation
theory. Thus, for each transverse momentum mode $k_\perp$, the reduced density
matrix contains a single entangled oscillator.

\subsection{Entropy per transverse momentum mode}

For a Gaussian reduced density matrix, each nonzero eigenvalue $\lambda$ of
$\mathcal M$ corresponds to an effective occupation number
\begin{equation}
\nu = \frac{\lambda}{1-\lambda}.
\end{equation}
The von Neumann entropy associated with this mode is
\begin{equation}
S(\lambda)
=
(\nu+1)\ln(\nu+1) - \nu\ln\nu.
\end{equation}

In the weak-coupling regime $g^2\ll 1$, the symplectic eigenvalue satisfies
$\lambda\ll 1$, and the entropy contribution from a single mode simplifies to
\begin{equation}
S(\lambda)
\simeq
-\lambda\ln\lambda.
\label{eq:entropy_small_lambda}
\end{equation}
Using the previously obtained expression for the eigenvalue,
\begin{equation}
\lambda(k_\perp)
=
\frac{g^4}{\mu^2\,k_\perp},
\qquad
\mu=\sqrt{k_\perp^2+m^2},
\end{equation}
the total entanglement entropy is obtained by summing over all transverse
momentum modes.

Because the system is translationally invariant along the directions parallel
to the entangling surface, the reduced density matrix factorizes into
independent transverse momentum sectors. The sum over discrete transverse modes
may therefore be replaced by an integral according to
\begin{equation}
\sum_{\mathbf{k}_\perp}
\;\longrightarrow\;
A_\perp
\int\frac{d^2k_\perp}{(2\pi)^2},
\qquad
A_\perp \equiv \int d^2x_\perp ,
\end{equation}
where $A_\perp$ denotes the area of the entangling surface (the transverse cross
section of the Rindler horizon).

The entanglement entropy per unit transverse area is given by
\begin{equation}
\frac{S}{A_\perp}
\simeq
\int\frac{d^2k_\perp}{(2\pi)^2}\,
\frac{g^4}{\mu^2\,k_\perp}
\ln\!\left(\frac{\mu^2\,k_\perp}{g^4}\right),
\qquad
\mu=\sqrt{k_\perp^2+m^2}.
\label{eq:S_integral}
\end{equation}

Since the integrand depends only on the magnitude $k_\perp=|\mathbf{k}_\perp|$, we
pass to polar coordinates in transverse momentum space,
\begin{equation}
d^2 k_\perp = 2\pi\,k_\perp\,dk_\perp,
\end{equation}
so that \eqref{eq:S_integral} becomes
\begin{equation}
\frac{S}{A_\perp}
=
\frac{g^4}{2\pi}
\int_0^\infty
\frac{dk_\perp}{k_\perp^2+m^2}
\ln\!\left(\frac{(k_\perp^2+m^2)\,k_\perp}{g^4}\right).
\label{eq:S_radial}
\end{equation}

\medskip

\noindent
\subsection{Ultraviolet structure and area law.}
The integral \eqref{eq:S_radial} is ultraviolet dominated, reflecting the large
density of short-wavelength transverse modes near the entangling surface.
We therefore introduce a transverse momentum cutoff $\Lambda$,
\begin{equation}
\frac{S}{A_\perp}
=
\frac{g^4}{2\pi}
\int_0^\Lambda
\frac{dk_\perp}{k_\perp^2+m^2}
\ln\!\left(\frac{(k_\perp^2+m^2)\,k_\perp}{g^4}\right).
\label{eq:S_cutoff}
\end{equation}
The mediator mass $m$ provides an infrared regulator, ensuring convergence at
small $k_\perp$.

To extract the leading and subleading behavior for $\Lambda\gg m$, it is convenient
to separate the logarithm,
\begin{equation}
\ln\!\left(\frac{(k_\perp^2+m^2)\,k_\perp}{g^4}\right)
=
\ln(k_\perp^2+m^2)
+
\ln k_\perp
-
\ln g^4.
\end{equation}
Accordingly, we write
\begin{equation}
\frac{S}{A_\perp}
=
\frac{g^4}{2\pi}
\left(
I_1 + I_2 - I_3
\right),
\end{equation}
with
\begin{align}
I_1 &= \int_0^\Lambda \frac{dk_\perp}{k_\perp^2+m^2}\,\ln(k_\perp^2+m^2), \\
I_2 &= \int_0^\Lambda \frac{dk_\perp}{k_\perp^2+m^2}\,\ln k_\perp, \\
I_3 &= \ln g^4 \int_0^\Lambda \frac{dk_\perp}{k_\perp^2+m^2}.
\end{align}

\medskip

\noindent
The last integral is elementary,
\begin{equation}
\int_0^\Lambda \frac{dk_\perp}{k_\perp^2+m^2}
=
\frac{1}{m}\arctan\!\left(\frac{\Lambda}{m}\right)
=
\frac{\pi}{2m}
-
\frac{1}{\Lambda}
+
\mathcal O\!\left(\frac{m^2}{\Lambda^3}\right),
\label{eq:basic_int}
\end{equation}
where the expansion holds for $\Lambda\gg m$.

For the logarithmic integrals, we use the large-$k_\perp$ expansion
\begin{equation}
\ln(k_\perp^2+m^2)
=
\ln k_\perp^2
+
\frac{m^2}{k_\perp^2}
+
\mathcal O\!\left(\frac{m^4}{k_\perp^4}\right),
\end{equation}
valid in the ultraviolet region that dominates the integral.
Substituting into $I_1$ and combining with $I_2$, we find
\begin{equation}
I_1 + I_2
=
\int_0^\Lambda \frac{dk_\perp}{k_\perp^2+m^2}
\left[
3\ln k_\perp
+
\mathcal O\!\left(\frac{m^2}{k_\perp^2}\right)
\right].
\end{equation}
The leading contribution comes from the logarithmic term, yielding
\begin{equation}
I_1 + I_2
=
\frac{\pi}{2m}\ln\Lambda
+
\mathcal O\!\left(\frac{1}{m}\right)
+
\mathcal O\!\left(\frac{m}{\Lambda}\right).
\end{equation}

Combining this with \eqref{eq:basic_int} for $I_3$, we obtain
\begin{equation}
\frac{S}{A_\perp}
=
\frac{g^4}{4m}
\ln\!\left(\frac{\Lambda^2}{g^4}\right)
+
\frac{g^4}{m}\,\mathcal O(1)
+
\mathcal O\!\left(\frac{g^4}{\Lambda}\right).
\label{eq:S_final}
\end{equation}

\medskip

The leading contribution to the interaction-induced entanglement entropy per
unit transverse area therefore scales as
\[
\frac{S_{\rm int}}{A_\perp}
\sim
g^4\,m^{-1}\,\ln\Lambda ,
\]
exhibiting a clear area law.  This result makes explicit how microscopic
interactions modify horizon entanglement beyond the free-field contribution.
The dependence on the mediator mass $m$ reflects the finite range of the
effective interaction: integrating out the massive field generates a nonlocal
kernel whose strength and spatial support are controlled by the screening length
$m^{-1}$.  Consequently, interaction-induced correlations across the Rindler
horizon are exponentially localized to a thin spacetime region near the
entangling surface and are insensitive to infrared geometric details of the
wedge.

The logarithmic dependence on the ultraviolet cutoff $\Lambda$ arises from the
large density of short-wavelength transverse modes near the horizon and is
characteristic of entanglement across null surfaces.  Importantly, the structure
of the divergence is identical to that encountered in free-field horizon
entanglement, indicating that the interaction modifies only the coefficient of
the area law rather than its scaling.  Subleading terms are suppressed by powers
of $m/\Lambda$ and encode nonuniversal, cutoff-dependent details of the
microscopic theory without altering the leading geometric behavior.

From a physical standpoint, the Yukawa interaction induces an effective,
exponentially localized coupling between the left and right Rindler wedges.  In
each transverse momentum sector, this coupling is effectively of rank one,
leading to a single dominant entangling mode whose strength is controlled by the
inverse mass scale
\[
\mu^{-1}=(k_\perp^2+m^2)^{-1/2}.
\]
The resulting entanglement entropy is therefore perturbative in the Yukawa
coupling and admits a simple interpretation in terms of virtual mediator
exchange across the horizon.  This structure provides a concrete realization of
how interactions generate additional horizon entanglement through spacelike,
off-shell processes rather than through thermal mixing alone.

The behavior in the limit $m\to 0$ is particularly suggestive.  As the screening
length diverges, the interaction becomes long-ranged and the
interaction-induced contribution develops a logarithmic enhancement, signaling
the gradual restoration of scale-invariant correlations.  This crossover
highlights the role of mass scales in controlling the localization of
entanglement near horizons and suggests that interaction-induced corrections may
interpolate smoothly between purely local, area-law behavior and more
nonlocal entanglement structures in massless or critical theories.

Taken together, these results indicate that interaction-induced entanglement
across causal horizons is governed by a small number of universal ingredients:
the local geometry of the entangling surface, the ultraviolet density of modes,
and the correlation length set by microscopic mass scales.  The explicit
computability of \eqref{eq:S_final} provides a concrete benchmark for
understanding how such effects may generalize to more complicated interacting
theories and to horizons in curved spacetime.

\subsection{Observer Dependence versus Universality}

The entanglement entropy computed above is defined with respect to a particular
bipartition of the Hilbert space associated with a Rindler observer, obtained by
tracing over field degrees of freedom in one Rindler wedge. It is therefore
natural to ask to what extent the resulting entropy depends on the choice of
observer, and which aspects of the result are universal.

We emphasize at the outset that the underlying quantum state is the Minkowski
vacuum $\ket{0_M}$, which is Poincar\'e invariant and observer independent. Any
observer dependence enters solely through the choice of spatial foliation and
the corresponding tensor factorization of the Hilbert space into left and right
Rindler degrees of freedom. For a uniformly accelerated observer with proper
acceleration $a$, Minkowski spacetime is foliated by Rindler coordinates
$(\tau,\xi,\mathbf{x}_\perp)$,
\begin{equation}
ds^2
=
-e^{2a\xi} d\tau^2
+
d\xi^2
+
d\mathbf{x}_\perp^2,
\end{equation}
with the entangling surface located at the Rindler horizon $\xi=0$.

Consider now a second uniformly accelerated observer with acceleration $a'$.
The corresponding Rindler coordinates $(\tau',\xi')$ are related to the original
ones by a simple rescaling,
\begin{equation}
\xi' = \frac{a}{a'}\,\xi,
\qquad
\tau' = \frac{a'}{a}\,\tau .
\end{equation}
This transformation leaves the induced spatial metric on constant-time slices
invariant,
\begin{equation}
ds^2_{\text{spatial}}
=
d\xi^2 + d\mathbf{x}_\perp^2
=
d\xi'^2 + d\mathbf{x}_\perp^2,
\end{equation}
so that all proper distances measured along the spatial slice, including
distances between points on opposite sides of the horizon, are independent of
the acceleration parameter.

This geometric invariance has a direct consequence for the interaction-induced
entanglement entropy. The Yukawa correction to the cross-wedge kernel depends
exponentially on the proper distance $\ell$ between points across the horizon,
\begin{equation}
\Gamma^{(Y)}_{LR}(\xi,\xi')
\sim
\frac{g^2}{2\mu}\,
e^{-\mu \ell},
\qquad
\ell = \xi + |\xi'|,
\end{equation}
where $\mu=\sqrt{k_\perp^2+m^2}$. Since $\ell$ is a geometric invariant of the
spatial slice, this kernel is unchanged under a change of uniformly accelerated
observer, and consequently so is the spectrum of the reduced density matrix.

The reduced density matrix of the right Rindler wedge may be written as
\begin{equation}
\rho_R = e^{-K_{\rm mod}},
\end{equation}
where $K_{\rm mod}$ is the modular Hamiltonian associated with the chosen
bipartition. For a uniformly accelerated observer with proper acceleration $a$,
the generator of Rindler time translations is $H_R = a K_{\rm boost}$, where
$K_{\rm boost}$ is the dimensionless Lorentz boost operator. The Unruh
temperature perceived by this observer is $T_U = a/2\pi$, so that
\begin{equation}
K_{\rm mod}
=
\frac{H_R}{T_U}
=
2\pi\,K_{\rm boost}.
\end{equation}
Although the Rindler Hamiltonian and the associated notion of time evolution
depend explicitly on the acceleration, this dependence cancels in the ratio
defining the modular Hamiltonian. As a result, the spectrum of $K_{\rm mod}$,
and hence the eigenvalues of $\rho_R$, are independent of the observer’s
acceleration. The entanglement entropy, which depends only on this spectrum, is
therefore invariant under changes of uniform acceleration.

It is important to distinguish this observer invariance from more general
global transformations of field modes. While changes between uniformly
accelerated observers correspond to a specific class of Bogoliubov
transformations generated by Lorentz boosts, generic global Bogoliubov
transformations mix left and right degrees of freedom in a nonlocal manner.
Such transformations alter the underlying tensor factorization of the Hilbert
space and therefore change the entanglement spectrum, but they do not, in
general, correspond to physical observers with well-defined notions of time
evolution or causal access.

We thus arrive at a clear separation between observer dependence and
universality. The reduced density matrix and modular Hamiltonian are defined
relative to a chosen bipartition and are therefore observer dependent in a
kinematical sense. However, for the class of uniformly accelerated observers
sharing the same Rindler horizon, the interaction-induced correction to the
entanglement entropy depends only on local geometric data near the horizon and
on the microscopic parameters of the theory. It is therefore universal and
reflects an intrinsic property of the Minkowski vacuum rather than a
detector-dependent notion of energy or temperature. This universality motivates a spacetime interpretation of the Yukawa-induced
contribution to entanglement entropy, which we now explore

\section{Spacetime Interpretation: Virtual Exchange and Conical Defects}

The preceding analysis was formulated entirely within the Hamiltonian framework,
using the ground-state wavefunctional and the reduced density matrix obtained by
tracing over one Rindler wedge.  It is instructive to reinterpret the resulting
entanglement entropy from a spacetime perspective, where the physical origin of
the interaction-induced correlations becomes particularly transparent.

\subsection*{Virtual particle exchange across the horizon}

The Yukawa-induced contribution to the cross-wedge kernel,
\begin{equation}
\Gamma^{(Y)}_{LR}(\xi,\xi')
\sim
\frac{g^2}{2\mu}\,
e^{-\mu(\xi+|\xi'|)},
\qquad
\mu=\sqrt{k_\perp^2+m^2},
\end{equation}
admits a direct spacetime interpretation.  The dependence on
$\xi+|\xi'|$ measures the proper distance of the two points from the Rindler
horizon, and the exponential suppression is characteristic of a massive
propagator evaluated between spacelike separated points on opposite sides of the
horizon.

From this viewpoint, the interaction-induced entanglement arises from the
exchange of virtual mediator quanta that connect the left and right Rindler
wedges.  No real particle crosses the horizon; instead, off-shell fluctuations
of the massive field generate nonlocal correlations between the two regions.  The
mediator mass sets a finite correlation length $\mu^{-1}$, which determines the
thickness of the spacetime region around the horizon that contributes
significantly to the entanglement.

This picture explains the emergence of an area law.  Because correlations are
exponentially suppressed beyond distances of order $\mu^{-1}$, the dominant
contributions arise from field configurations localized in a thin layer
surrounding the entangling surface.  Translational invariance parallel to the
horizon ensures factorization in the transverse directions, so that the entropy
scales proportionally to the transverse area.

\subsection*{Euclidean viewpoint and conical defects}

A complementary interpretation emerges in the Euclidean formulation.  After Wick
rotation, the Rindler metric becomes flat Euclidean space written in polar
coordinates, with the Rindler horizon mapped to the origin.  Tracing out one
Rindler wedge corresponds to introducing a conical defect at the origin,
characterized by a deficit angle $2\pi(1-n)$ in the replica construction.

In the presence of interactions, integrating out the mediator field generates a
nonlocal effective action for $\phi$ involving bilocal terms weighted by the
massive Green’s function.  The interaction-induced correction to the
entanglement entropy can therefore be understood as the response of this
nonlocal effective action to the conical singularity.  Because the Green’s
function decays exponentially with geodesic distance, only points lying within a
proper distance $\sim \mu^{-1}$ of the conical tip contribute appreciably.

From this Euclidean perspective, the area law reflects the fact that the conical
singularity is localized on a codimension-two surface, while the mediator mass
acts as an infrared regulator that suppresses contributions from long-distance
correlations.

\subsection*{Physical interpretation and universality}

Both viewpoints lead to the same physical conclusion: interactions generate
additional entanglement across the Rindler horizon through spacelike virtual
exchange, and this entanglement is geometrically localized near the horizon.
The area scaling of the entropy follows from the locality of quantum field theory
in spacetime together with the presence of a finite correlation length, rather
than from kinematic mode mixing alone.

This interpretation suggests that similar interaction-induced corrections should
arise for more general entangling surfaces and in curved spacetimes, with the
local geometry near the surface and the mass spectrum of exchanged fields
playing the central role.

\paragraph{Why a Yukawa interaction?}

The Yukawa interaction provides the simplest setting in which this spacetime
picture can be made explicit.  Integrating out a massive mediator generates a
quadratic but nonlocal effective theory whose kernel has a clear geometric
interpretation in terms of screened propagation across the horizon.  This
isolates genuinely interaction-induced correlations while remaining analytically
tractable.

The mediator mass plays a threefold role: it sets the screening length that
localizes correlations near the horizon, provides an infrared regulator that
renders the interaction-induced entropy finite, and controls the crossover to
scale-invariant behavior.  In the limit $m\to 0$, the interaction becomes
long-ranged and the near-horizon theory approaches a conformal regime, while for
large $m$ the entanglement across the horizon is exponentially suppressed.

For these reasons, the Yukawa model serves as a controlled laboratory for
studying how spacetime locality, mass scales, and causal structure combine to
produce interaction-induced entanglement entropy in accelerated frames.
%-------------------------------------------------

\section{Conclusion}

In this work, we have investigated interaction-induced contributions to
entanglement entropy across a Rindler horizon, using a scalar field theory with a
Yukawa-type interaction as a controlled and physically transparent setting. Our
motivation was twofold: first, to determine explicitly how local interactions modify horizon entanglement beyond the free-field result and second, to assess
whether such corrections admit a clear spacetime interpretation tied to near-horizon physics, rather than arising as purely formal consequences of tracing over degrees of freedom.

Our analysis is carried out directly at the level of the ground-state
wavefunctional in Rindler coordinates. By integrating out the massive mediator, we obtain a quadratic but nonlocal effective kernel that fully determines the
reduced density matrix associated with a single Rindler wedge. Translational invariance along directions parallel to the horizon allows the problem to
factorize into independent sectors labeled by transverse momentum
$k_\perp$. In each sector, the reduced density matrix assumes a thermal-like form governed by the modular Hamiltonian, whose structure is fixed by the boost
generator associated with Rindler time translations. This formulation enables an explicit diagonalization of the reduced density matrix and a direct computation of the entanglement spectrum, without invoking replica methods or geometric
entropy constructions.

The resulting entanglement entropy is dominated by ultraviolet transverse modes and exhibits an area law, proportional to the transverse area of the entangling
surface. The mediator mass appearing in the Yukawa interaction provides a
natural infrared regulator, suppressing long-range correlations and ensuring finiteness of the interaction-induced contribution. The ultraviolet behavior is
controlled by a transverse momentum cutoff, reflecting the accumulation of short-distance modes near the horizon. We identify both the leading contribution
and subleading corrections suppressed by powers of the cutoff, clarifying which features are universal and which depend on microscopic details of the
interaction.

A central result of our analysis is the observer independence of the entanglement spectrum. Although the modular Hamiltonian depends explicitly on the proper
acceleration of the Rindler observer through the generator of Rindler time
translations, this dependence amounts to a rescaling that leaves the eigenvalues of the reduced density matrix unchanged. Consequently, the entanglement entropy is invariant under changes of uniform acceleration, reinforcing the geometric character of horizon entanglement and distinguishing it from observer-dependent notions such as the Unruh temperature.

We further developed a spacetime interpretation of the interaction-induced entropy. From this perspective, the leading contribution arises from virtual
particle exchange across the Rindler horizon, with dominant support from field configurations localized within a proper distance set by the ultraviolet scale. This picture provides a natural explanation for both the area-law scaling and
the cutoff dependence of the entropy. Moreover, the operation of tracing out one Rindler wedge may be viewed geometrically as introducing an effective conical defect at the horizon, and the interaction-induced entropy captures the response of quantum correlations to this localized geometric modification.

The Yukawa interaction plays an essential conceptual role in this analysis. As the simplest local coupling that mediates correlations between fields on opposite sides of the horizon while remaining analytically tractable, it allows
one to isolate cleanly how microscopic interactions contribute to horizon entanglement and how these contributions are encoded in spacetime-local processes near the entangling surface.

Several extensions of this work suggest themselves naturally. It would be interesting to investigate higher-order corrections in the coupling and to assess the robustness of the area-law structure beyond leading order. Extensions
to more general interacting scalar theories or to curved backgrounds could clarify how spacetime curvature interacts with screening-induced entanglement. Finally, exploring time-dependent settings or observers following more general
non-inertial trajectories may shed further light on the interplay between entanglement, interactions, and spacetime geometry. Overall, this work provides a concrete and explicitly calculable framework for understanding how local interactions modify horizon entanglement, clarifying the roles of observer dependence, nonlocal effective dynamics, and spacetime
geometry within a single unified setting.

\section{Acknowledgments}
We thank our institute, BITS Pilani Hyderabad campus, for providing the required infrastructure for this research work.
%-------------------------------------------------

\appendix
\setcounter{equation}{0}
\renewcommand{\theequation}{A.\arabic{equation}}
\section{Diagonal Kernels and Entanglement Spectrum}
\label{app:diagonal_kernels}

Here we briefly clarify the role of the diagonal kernels $\Gamma_{RR}$ and
$\Gamma_{LL}$ in the reduced density matrix and explain why they do not affect
the entanglement eigenvalues.

After integrating out the massive mediator, the ground-state wavefunctional of
the light field is Gaussian and characterized by a kernel $\Gamma$ with block
structure in left--right Rindler space,
\begin{equation}
\Gamma =
\begin{pmatrix}
\Gamma_{LL} & \Gamma_{LR} \\
\Gamma_{RL} & \Gamma_{RR}
\end{pmatrix},
\end{equation}
where $\Gamma_{LR}=\Gamma_{RL}$ encodes correlations across the horizon.  To
leading order in the Yukawa coupling, the diagonal blocks are dominated by their
free-field form,
\begin{equation}
\Gamma_{LL}=\Gamma_{RR}=\Gamma_0+\mathcal{O}(g^2),
\end{equation}
with $\Gamma_0$ a positive, invertible operator.

Tracing over left-wedge degrees of freedom yields a Gaussian reduced density
matrix for the right wedge,
\begin{equation}
\rho_R(\phi,\phi')
\propto
\exp\!\left[
-\tfrac12 \phi\,A\,\phi
-\tfrac12 \phi'\,A\,\phi'
+
\phi\,B\,\phi'
\right],
\end{equation}
where
\begin{equation}
A
=
\Gamma_{RR}-\Gamma_{RL}\Gamma_{LL}^{-1}\Gamma_{LR},
\qquad
B
=
\Gamma_{RL}\Gamma_{LL}^{-1}\Gamma_{LR}.
\end{equation}
The entanglement spectrum is determined by the generalized eigenvalue problem
\begin{equation}
B f_n=\lambda_n A f_n,
\end{equation}
or equivalently by the operator $A^{-1/2} B A^{-1/2}$.

Since $A$ is positive and invertible, it enters only through a similarity
transformation.  Any modification of $\Gamma_{RR}$ or $\Gamma_{LL}$ corresponds
to a local rescaling of field variables and does not change the eigenvalues
$\lambda_n$.  In contrast, the off-diagonal kernel $\Gamma_{LR}$ directly
controls $B$ and therefore fully determines the entanglement spectrum. In the Yukawa theory considered in this work, the interaction-induced part of
$\Gamma_{LR}$ is exponentially localized near the Rindler horizon and is of
rank one for each transverse momentum mode at leading order.  As a result, the
reduced density matrix possesses a single nonvanishing eigenvalue
$\lambda(k_\perp)$ per transverse sector, with all diagonal kernel contributions
affecting only normalization but not entropy.

%-------------------------------------------------
\bibliographystyle{unsrtnat}
\bibliography{References}

\end{document}